\documentclass[11pt,a4paper]{article}
\usepackage{a4wide}
\usepackage{amsmath}
\usepackage{url}
\usepackage{epsfig}
\usepackage{psfrag}
\usepackage{amssymb}
\usepackage{amsmath}

\usepackage{authblk}
\usepackage{wrapfig}
\usepackage{algorithm}
\usepackage{algorithmic}
\usepackage{paralist}
\usepackage{xcolor}
\usepackage{xfrac}
\usepackage{comment}
\usepackage[colorlinks=true,linkcolor=blue,citecolor=blue,urlcolor=blue]{hyperref}
\usepackage{cleveref}

\usepackage{mathptmx}
\usepackage[scaled]{helvet}   
\usepackage{courier}

\usepackage[T1]{fontenc}
\usepackage[utf8]{inputenc}

\newcommand{\qed}]
\newcommand{\witness}{(X_0, X_1, X_2)}
\newcommand{\sset}[1]{\{ #1 \}}
\newcommand{\low}{\mathit{low}}
\newcommand{\high}{\mathit{high}}
\newcommand{\R}{\mathbb{R}}

\newcommand{\PR}{\operatorname{PR}}

\newcommand{\calY}{\mathcal Y}

\newtheorem{theorem}{Theorem}
\newenvironment{proof}[1][Proof]{\textbf{#1:}\ }{\hfill $\blacksquare$}
\newtheorem{lemma}{Lemma}
\newtheorem{claim}{Claim}
\newcommand{\abs}[1]{| #1 |}

\begin{document}

\title{EFX Exists for Three Agents with One General Valuation and Two Compatible Valuations}
\author[1]{Eyad Alkassar \thanks{eyad.alkassar@gmail.com}}
\author[2]{Arash Ashuri\thanks{aashuri@mpi-inf.mpg.de}}
\author[1]{Mahmoud Fouz \thanks{mahmoudfouz@gmail.com}}
\author[2]{Kurt Mehlhorn \thanks{mehlhorn@mpi-inf.mpg.de}}

\affil[1]{Independent Researcher}
\affil[2]{Max Planck Institute for Informatics \& Saarland Informatics Campus, Saarbrücken, Germany}

\maketitle

\begin{abstract} An allocation of a set of indivisible goods to a set of agents is EFX (envy-free up to any good) if no agent envies any other agent after the removal of any good from the other agent's bundle. EFX-allocations exist for three agents, two of which have general monotone valuations and one of which has an MMS-valuation. We show that EFX-allocations exist for three agents, two of which have a compatible pair of valuations, and one of which has a general monotone valuation. A monotone valuation and an MMS-valuation form a compatible pair.  \end{abstract}

\tableofcontents

\section{Introduction}

There are three agents $0$, $1$, and $2$, and a set $M$ of items. For an agent $i$, $v_i: 2^M \rightarrow \R_{\ge 0}$ is its valuation. The valuations are assumed to be monotone, i.e., if $A \subseteq B$, then $v_i(A) \le v_i(B)$. 

    A \emph{one-deletion} of a set is the set minus one of its elements. For a set $B$ and an item $g$, we write $B \setminus g$ and $B \cup g$ for $B \setminus \{g\}$ and $B \cup \{g\}$, respectively. For an agent $i$ and a bundle $B$ of goods, $\rho_i(B) = \max_{x \in B} v_i(B \setminus x)$ is the maximum value of a one-deletion of $B$; $\rho_i(\emptyset) = 0$. A \emph{partial allocation} is a partition of a subset of $M$ into three bundles, one for each agent. In an allocation, all items of $M$ are allocated. We write a (partial) allocation as $(X_0,X_1,X_2)$, where $X_i$ denotes the set of items allocated to agent $i$.

    An agent $i$ with a bundle $B$ is \emph{EFX (envy-free up to any good) toward} a bundle $C$ if $\rho_i(C) \le v_i(B)$; otherwise, $i$ with $B$ \emph{strongly envies} $C$. In an allocation, agent $i$ is EFX toward agent $j$ (or agent $i$ does not strongly envy $X_j$ or agent $j$) if $i$ with $X_i$ is EFX toward $X_j$. An allocation is EFX if all agents are pairwise EFX, and it is EFX for agent $i$ if $i$ is EFX toward all other agents. We use these notions also for allocations of a subset of $M$ to two of the agents.

    For an agent $i$ and a partition $P = (P_1,P_2)$ of a set of items into two parts, let $\high_i(P) = \max\{v_i(P_1),v_i(P_2)\}$ and $\low_i(P) = \min\{v_i(P_1), v_i(P_2)\}$. A valuation $v_i$ is \emph{MMS-feasible} if for any subset of goods and any two partitions $P = (P_1,P_2)$ and $Q = (Q_1,Q_2)$ of this subset, we have $\low_i(P) \le \high_i(Q)$.
    It is known that an EFX allocation for three agents exists (\cite{AACGMM22}), if one of the agents has an MMS-feasible valuation.

Let $i$ and $j$ be agents. The agents form a \emph{compatible pair} if for any subset of items and any two partitions $P= (P_1,P_2)$ and $Q =(Q_1,Q_2)$ of it,
$$ \text{either } \high_i(P) \ge \low_i(Q) \text{ or } \high_j(Q) \ge\low_j(P).$$
In other words, it cannot be the case that
$$\high_i(P) < \low_i(Q) \text{ and } \high_j(Q) < \low_j(P),$$
i.e., $i$ ranks both parts of $Q$ better than both parts of $P$ and $j$ ranks both parts of $P$ better than both parts of $Q$. If $i$ has an MMS-feasible valuation and $j$ has an arbitrary monotone valuation, $i$ and $j$ form a compatible pair. Indeed, we have $\low_i(P) \le \high_i(Q)$ and, by reversing the roles of $P$ and $Q$, $\low_i(Q) \le \high_i(P)$ and hence the first disjunct in the compatible pair definition holds. Compatibility is more general than MMS-feasibility. For example, two agents with identical monotone valuations form a compatible pair.\footnote{A more substantial example is as follows: We have two pairs of shoes, a blue pair and a red pair. Agent $i$ has a regular taste; they want either a blue pair or a red pair, and a pair consisting of shoes of different colors has no value to $i$. Agent $j$ is a clown with a very particular taste; $j$ values a pair consisting of a red left and a blue right shoe the highest, followed by the standard pairs, followed by the pair consisting of a blue left and a red right shoe. Let $P$ be the partition into two standard pairs of shoes, and let $Q$ be the partition into two pairs of unequal color. Then $i$ values both parts of $P$ higher than both parts of $Q$, and $j$ values one pair higher than both parts of $P$ and one pair lower.  }

    In this note, we extend the result of~\cite{AACGMM22} to compatible pairs and prove:

\begin{theorem}\label{thm:main}
    EFX exists for three agents whenever two agents form a compatible pair and the third agent has a general monotone valuation.
\end{theorem}

We give two proofs of the theorem in Sections~\ref{proof1} and~\ref{proof2}, respectively. The second proof extends the proof in~\cite{AACGMM22} to compatible pairs; the first proof was found with the help of an AI   and precedes the second proof. See Section \ref{methods and disclosure} for details.


\section{Related Work}\label{sec:related}
    
    EFX was introduced by Caragiannis et al.~\cite{CKMPSW19} (see also~\cite{GMT14}). It strengthens envy-freeness up to one good (EF1)~\cite{Bud11}, for which allocations always exist and can be computed in polynomial time~\cite{LMMS04}. 

\paragraph{Existence for special instances.} 
    Plaut and Roughgarden~\cite{PR20} showed that EFX allocations exist for two agents with general monotone valuations and for any number of agents with identical general monotone valuations. The two-agent result is a cut-and-choose argument; \cref{lem:two-cutter} is a cut-and-choose argument in which both agents may cut. The PR repair used in \cref{proof2} is their procedure. For three agents, Chaudhury, Garg, and Mehlhorn~\cite{CGM24} proved that EFX allocations exist for additive valuations, and Berger et al.~\cite{BCFF22} extended this to nice-cancelable valuations. Akrami et al.~\cite{AACGMM22} gave a much simpler proof and showed that EFX allocations exist for three agents if one agent has an MMS-feasible valuation and the other two have arbitrary monotone valuations. MMS-feasible valuations strictly generalize cancelable valuations and include additive, budget-additive, and unit-demand valuations. 
    Although many existence results for EFX under additive valuations have been extended to cancelable valuations, the two classes are distinct even at the ordinal level. Akrami et al. \cite{AACMS26} showed that there exist cancelable valuations whose preferences over bundles cannot be represented by any additive valuation.
    
    Mahara~\cite{Mahara24} showed that EFX allocations exist for any number of agents if every agent has one of two general monotone valuations, and if the number of goods is at most $n + 3$, where $n$ is the number of agents. For additive valuations, EFX allocations exist for any number of agents with at most three distinct valuations~\cite{HGNV25}, if every good has one of two values~\cite{ABFHV21}, and for four agents and at most nine goods~\cite{AFM26}. EFX allocations also exist for graphical valuations, where agents are the vertices of a graph, goods are its edges, and every agent values only the edges incident to it~\cite{CFKS23}; this was recently extended to multigraphs with cancelable valuations~\cite{AAMMS26} and to triangle-free multigraphs with monotone valuations \cite{AAMM26}.
    
\paragraph{Relaxations.} 
    EFX allocations exist if some goods may remain unallocated: at most $n-1$ goods such that no agent envies the set of unallocated goods~\cite{CKMS21}, and at most $n-2$ goods for additive~\cite{BCFF22} and for general monotone valuations~\cite{Mahara24}. For additive valuations, $(\phi - 1)$-EFX allocations exist, where $\phi$ is the golden ratio~\cite{AMN20}. EF2X allocations, where envy disappears after the removal of any two goods, exist for four agents with cancelable valuations~\cite{AGS25}. On the computational side, the PR repair can be made polynomial for cancelable valuations~\cite{AACMS26}, whereas computing an EFX allocation for two agents with submodular valuations is PLS-complete~\cite{GHH25}.

   Several alternative relaxations of EFX have also been proposed, including EFL \cite{BBMN18}, EEFX and MXS \cite{CGRSV22}, which were later extended to monotone valuations by \cite{HANR25}. Related work has also examined the compatibility of different fairness guarantees, such as the coexistence of $2/3$-MMS with EF1 \cite{AR25}, the combination of MXS and EFL \cite{AG25}, and allocations satisfying RMMS, EEFX, and EFL \cite{AFMMR26}.
   Finally, this was extended to $2/3$-APS, $1-\epsilon$-MXS, and EFL computable in polynomial time in the number of agents and goods and $1/\epsilon$ \cite{AAM26}. 

\paragraph{Non-existence.} 
    Akrami et al. \cite{AMMMSSW26} showed that EFX allocations need not exist for three agents with weighted-coverage (and hence submodular) valuations over eight goods. Their construction extends to $n\geq 3$ agents and $m\geq n+5$ goods. For three agents, eight goods are necessary, since every instance with at most seven goods and monotone valuations admits an EFX allocation. They further showed that $\alpha$-EFX allocations need not exist for $\alpha>2^{-1/6}\approx 0.891$ under subadditive valuations or for $\alpha>0.958$ under submodular valuations.

\section{Proof One}\label{proof1}

\emph{From now on, we assume that agents 1 and 2 form a compatible pair.} The agent $0$ plays a special role in the construction of the EFX allocation. A set $X_0$ is \emph{feasible} for $0$ or simply \emph{feasible} if there is an  allocation $(X_0,X_1,X_2)$ such that agents 1 and 2 do not strongly envy any of the other two agents, but $0$ might strongly envy any of the other two agents. Informally, a set $X_0$ is feasible for agent 0 if the remaining items can be assigned to agents 1 and 2 such that they are both happy in the EFX sense. Note that feasible means feasible for $0$. Also note that feasible means not too heavy; it does not mean sufficiently heavy. We call the allocation a \emph{completion} of $X_0$. If $X_0$ is feasible any subset $W$ of $X_0$ is also feasible. Indeed, let $R = X_0 \setminus W$ and apply Lemma~\ref{Extending} below.

\subsection{Preliminaries}\label{Preliminaries}

\begin{lemma}[Cut-and-Choose with Two Thresholds] \label{Split with two bounds}\label{lem:two-cutter} Let $i$ and $j$ be a compatible pair of agents, let $S$ be a set of items, let $P$ and $Q$ be two partitions of $S$, and let $\mu_i \le \low_i(P)$ and $\mu_j \le \low_j(Q)$. Then there is an EFX allocation of $S$ to $i$ and $j$ with value at least $\mu_i$ for $i$ and value at least $\mu_j$ for $j$. \end{lemma}
\begin{proof} Let $i$ balance $P$, i.e., as long as $i$ with the less valuable part of the partition strongly envies the more valuable part, move an item whose removal decreases the value of the more valuable part the least from the more valuable part to the less valuable part. This does not decrease the minimum of the two values, and the process terminates~\cite{PR20}. Let $P^*$ be the resulting partition. Let $j$ pick the part more valuable to them. If it has value at least $\mu_j$, we are done. 

Otherwise, $\high_j(P^*) < \mu_j$ and $\low_i(P^*) \ge \low_i(P) \ge \mu_i$. 
Let $j$ balance $Q$, call the resulting partition $Q^*$, and let $i$ pick. If it receives a bundle of value at least $\mu_i$, we are done. Otherwise, $\high_i(Q^*) < \mu_i$ and $\low_j(Q^*) \ge \low_j(Q) \ge \mu_j$. 

  If both splits fail, $\low_i(P^*) \ge \mu_i > \high_i(Q^*)$ and $\low_j(Q^*) \ge \mu_j >  \high_j(P^*)$, witnessing the incompatibility of $i$ and $j$.
\end{proof}


\begin{lemma}[Extending a Partial Two-Agent EFX Allocation] \label{extending}\label{Extending} Let agents 1 and 2 form a compatible pair, let $(X_0, X_1, X_2)$ be a partial allocation, in which agents 1 and 2 do not strongly envy each other, and let $R$ be a set of unassigned goods. Then there is a split of $X_1 \cup X_2 \cup R$ into two sets $Z_1$ and $Z_2$ such that for $i \in \{1,2\}$, agent $i$ with $Z_i$ does not strongly envy the other. Moreover, if $i$ with $X_i$ did not strongly envy $X_0$, the same holds true for $Z_i$.
\end{lemma}
\begin{proof} Consider the partitions $(X_1 \cup R,X_2)$ and $(X_1, X_2 \cup R)$. If the first fails to be EFX, we must have $v_2(X_2) < \rho_2(X_1 \cup R) \le v_2(X_1 \cup R)$. If the second fails to be EFX, we have $v_1(X_1) < \rho_1(X_2\cup R) \le v_1(X_2 \cup R)$. Thus, if both fail, Lemma~\ref{Split with two bounds} guarantees a partition of $X_1 \cup X_2 \cup R$ that has value at least $v_1(X_1)$ to $1$ and at least $v_2(X_2)$ to $2$. Since the new bundles for $1$ and $2$ have at least the value of the old bundles, no strong envy toward $X_0$ can be created. \end{proof}



\begin{lemma}[Feasibility Criterion]\label{Feasibility Criterion} Assume agents 1 and 2 form a compatible pair and do not strongly envy each other in the allocation $(X_0,X_1,X_2)$. If agent 1 strictly prefers $X_0$ to $X_1$ or agent 2 strictly prefers $X_0$ to $X_2$, then $X_1$ or $X_2$ is feasible.\end{lemma}
\begin{proof} Assume that either 1 or 2 strictly prefers $X_0$ over their set. Let $W$ be an inclusion-minimal subset of $X_0$ such that either $1$ or $2$ strictly prefers $W$ to their set. Then $\rho_i(W) \le v_i(X_i)$ for $i \in \sset{1,2}$. Say, $1$ strictly prefers $W$ over $X_1$, the other case being symmetric. Then $v_1(X_1) < v_1(W)$. Consider the partial allocation $(X_1, W, X_2)$, i.e., 0 receives $X_1$, 1 receives $W$, and 2 keeps $X_2$.

Then $1$ with $W$ does not strongly envy $X_2$, since $1$ prefers $W$ over $X_1$ and did not strongly envy $X_2$ with $X_1$. Also, $1$ does not strongly envy $X_1$, since it prefers $W$ over $X_1$.

$2$ with $X_2$ does not strongly envy $W$ since $\rho_2(W) \le v_2(X_2)$. Moreover, it clearly does not strongly envy $X_1$.

An application of Lemma~\ref{extending} with $R = X_0 \setminus W$ completes the proof. \end{proof}

\subsection{The Proof of the Theorem}\label{proof of main theorem}

\begin{proof} We prove the main theorem. Consider a most valuable feasible set $X_0$. Most valuable means most valuable to $0$; among the most valuable feasible sets, we choose an inclusion-maximal one, and hence, by monotonicity, $X_0$ is an inclusion-maximal feasible set. $X_0$ exists since the empty set and all singletons are feasible. By Lemma~\ref{construction of a witness} below, there is a completion $(X_0,X_1,X_2)$ of $X_0$ such that all one-deletions of $X_1$ and $X_2$ are feasible. Such a completion is an EFX allocation, since $X_0$ is a most valuable feasible set and hence at least as valuable as all one-deletions of $X_1$ and $X_2$. Thus agent 0 with $X_0$ does not strongly envy either $X_1$ or $X_2$. 
\end{proof}

How do we arrive at such a completion? We start with a completion $\witness$ of $X_0$. If all one-deletions of $X_1$ and $X_2$ are feasible, we are done. Otherwise, the following Lemma shows how to modify the completion such that all one-deletions of the bundles of agents 1 and 2 become feasible. The modification does not change $X_0$. In a nutshell, the construction is as follows. Assume w.l.o.g.~that $X_1$ has an infeasible one-deletion. Then let $X_1'$ be a minimal (with respect to set inclusion) infeasible subset of such a one-deletion and let $X_2' = X_2 \cup(X_1 \setminus X_1')$. Then one first shows that $(X_0, X_1', X_2')$ is a completion of $X_0$; this is the core of the proof. One then turns to the one-deletions of $X_1'$ and $X_2'$. By definition, every one-deletion of $X_1'$ is feasible. If every one-deletion of $X_2'$ is feasible, we are done. Otherwise, we repeat the construction with $X_2'$ playing the role of $X_1$, i.e., we let $I$ be an inclusion-minimal infeasible subset of an infeasible one-deletion of $X_2'$ and conclude that $(X_0, X_1' \cup (X_2' \setminus I), I)$ is a completion of $X_0$. Finally, consider $(X_0 \cup (X_2' \setminus I), X_1', I)$. In this allocation, 1 with $X_1'$ and 2 with $I$ do not strongly envy each other, and the set allocated to $0$ is infeasible since it is a proper superset of $X_0$. Thus, either 1 or 2 must strongly envy agent 0 in this allocation. Thus, either $I$ or $X_1'$ is feasible by Lemma~\ref{Feasibility Criterion}, but both are infeasible. This contradiction completes the argument.

\begin{lemma}[Construction of a Completion]\label{construction of a witness} Let $X_0$ be an inclusion-maximal feasible set,\footnote{Being most valuable is only needed in the proof of Theorem~\ref{thm:main} above. Maximality suffices for the rest of the proof.} let $\witness$ be a completion of $X_0$, and assume that there is an infeasible one-deletion of either $X_1$ or $X_2$. By interchanging the roles of agents 1 and 2, if necessary, we may assume that there is an infeasible one-deletion of $X_1$. Let $X_1'$ be a minimal (with respect to set inclusion) infeasible subset of such an infeasible one-deletion, let $X_2' =  X_2 \cup (X_1 \setminus X_1')$, and consider
\[        (X_0, X_1', X_2'). \]
The allocation is a completion of $X_0$. Every one-deletion of $X_1'$ and $X_2'$ is feasible. \end{lemma}
\begin{proof} Note that $X_1 \setminus X_1'$ is non-empty. By definition, $X_1'$ is infeasible and has feasible one-deletions. 

We first show that $2$ with $X_2'$ is EFX toward the other two sets. In fact, we show the stronger claims
\begin{equation}\label{H is good for a}     \rho_2(X_2') \ge v_2(X_1') \quad \text{and}\quad  v_2(X_2')  \ge \rho_2(X_0). \end{equation}
  We have 
  \[ \rho_2(X_2') \ge v_2(X_2) \ge \rho_2(X_1) \ge v_2(X_1'). \]
The first inequality holds since $X_2$ is a proper subset of $X_2'$, the second holds since $2$ with $X_2$ is EFX toward $X_1$, and the third holds since $X_1'$ is a proper subset of $X_1$. The second inequality in (\ref{H is good for a}) holds since $2$ with $X_2$ and hence with $X_2'$ is EFX toward $X_0$.\medskip

We next show that $1$ with $X_1'$ is EFX toward the other two sets. In fact, we show the stronger claims
\begin{equation}\label{T is good for d} \rho_1(X_1') > v_1(X_0) \quad\text{and}\quad \rho_1(X_1') > \rho_1(X_2'). \end{equation}
Let $N = X_0 \cup X_2' = X_0 \cup X_2 \cup (X_1 \setminus X_1') = M \setminus X_1'$, and let $r = \rho_1(X_1')$. We establish four claims from which (\ref{T is good for d}) follows easily. 
  
  \begin{claim}\label{a-split} $N$ has a partition whose two parts have value at least $v_2(X_1')$ to $2$. \end{claim}
  \begin{proof} Let $X_2''$ be an inclusion-minimal subset of $X_2'$ with $v_2(X_2'') \ge v_2(X_1')$. Then $X_2''$ is a proper subset of $X_2'$ since $\rho_2(X_2') \ge v_2(X_1')$ by (\ref{H is good for a}). Also, $\rho_2(X_2'') < v_2(X_1')$ by minimality of $X_2''$. 
Let $X_0'' = N \setminus X_2'' = X_0 \cup (X_2' \setminus X_2'')$. We show $v_2(X_0'') \ge v_2(X_1')$.

Assume $v_2(X_0'') < v_2(X_1')$. $X_0''$ is a proper superset of $X_0$ and hence infeasible. The three sets $X_2''$, $X_0''$, and $X_1'$ partition $M$, since $X_0''$ and $X_2''$ partition $N$, and $N = M \setminus X_1'$. Agent $2$ with $X_1'$ is EFX toward the other two bundles since $v_2(X_1') > \rho_2(X_2'')$ by the previous paragraph and $v_2(X_1') > v_2(X_0'')$ by assumption, and $2$ with $X_2''$ is EFX toward the other two bundles since $v_2(X_2'') \ge v_2(X_1') \ge v_2(X_0'')$. Let $1$ pick its favorite bundle. If $1$ picks $X_0''$ or $X_1'$, give the other of the two sets to $0$ and give $X_2''$ to $2$. If $1$ picks $X_2''$, give $X_1'$ to $2$ and give $X_0''$ to $0$. 
So $0$ gets $X_0''$ or $X_1'$. In either case, agents 1 and 2 do not strongly envy any of the two other agents, and hence the set given to 0 is feasible. But $X_0''$ is a proper superset of $X_0$ and hence infeasible, and $X_1'$ is infeasible by definition. \end{proof} 

\begin{claim}\label{claim 2} No partition of $N$ has both parts worth at least $r$ to $1$. Recall $r = \rho_1(X_1')$. \end{claim}
\begin{proof} Otherwise, apply Lemma~\ref{Split with two bounds} to $N$ with threshold $r$ for $1$ and threshold $v_2(X_1')$ for $2$. The required split exists for $1$ by assumption of the claim and exists for $2$ by Claim~\ref{a-split}. We obtain a partition of $N$ for $1$ and $2$, in which $1$ and $2$ do not strongly envy each other, $2$'s bundle has value at least $v_2(X_1')$ to $2$, and $1$'s bundle has value at least $r$ to $1$. Hence, the agents with their bundles are EFX toward $X_1'$. So the partition is a completion of $X_1'$, a contradiction to the infeasibility of $X_1'$. \end{proof} 

\begin{claim} $v_1(X_0) < r$, i.e., the value of $X_0$ to $1$ is less than $r$. \end{claim}
\begin{proof} Suppose, $v_1(X_0) \ge r$. Application of the previous claim to the partition $(X_0,X_2')$ of $N$ gives $v_1(X_2') < r$. Hence, $(X_1',X_0,X_2')$ is a completion of $X_1'$, a contradiction to the infeasibility of $X_1'$. Indeed, $1$ with $X_0$ is EFX toward $X_2'$, since $v_1(X_2') < r \le v_1(X_0)$, and toward $X_1'$, since $\rho_1(X_1') = r \le v_1(X_0)$, and $2$ with $X_2'$ is EFX toward $X_0$ and $X_1'$ by  (\ref{H is good for a}). \end{proof}

  \begin{claim}  $\rho_1(X_2') < r$, i.e., the value to $1$ of any one-deletion of $X_2'$ is less than $r$.  \end{claim}
  \begin{proof} Assume otherwise. Then some one-deletion of $X_2'$ has value at least $r$ to $1$. Let $U$ be an inclusion-minimal subset of this one-deletion with $v_1(U) \ge r$. Then $U$ is a proper subset of $X_2'$ and  $\rho_1(U) < r$. Hence, $V = N \setminus U = X_0 \cup (X_2' \setminus U)$ is a proper superset of $X_0$ and hence infeasible. Since $v_1(U) \ge r$, we have $v_1(V) < r$ by Claim~\ref{claim 2}.  The three sets $U$, $V$, and $X_1'$ partition $M$. Agent $1$ is EFX toward the other two bundles whether it holds $X_1'$ or $U$, since $v_1(X_1') \ge \rho_1(X_1') \ge r$, $v_1(U) \ge r$, $r \ge \rho_1(U)$, and $r \ge v_1(V)$. Let $2$ pick its favorite bundle. If $2$ picks $X_1'$ or $U$, give the other of the two sets to $1$, and give $V$ to $0$. If $2$ picks $V$, give $U$ to $1$ and give $X_1'$ to $0$. So, $0$ receives either $V$ or $X_1'$. In either case, agents 1 and 2 do not strongly envy any of the two other agents, and hence the set given to 0 is feasible. But, both are not.  \end{proof}
  
  We have now established that $2$ with $X_2'$ is EFX toward the other two bundles and $1$ with $X_1'$ is EFX toward the other two bundles. So, $(X_0,X_1',X_2')$ is a completion of $X_0$.

  It remains to show that every one-deletion of $X_2'$ is feasible. Assume otherwise, and let $I$ be a minimal (with respect to set inclusion) infeasible subset of an infeasible one-deletion of $X_2'$. Then, $I$ is a proper subset of $X_2'$, and, by the first part of the proof with the roles of $2$ and $1$ interchanged, 
  \[                    (X_0,X_1' \cup (X_2' \setminus I), I) \]
  is a completion of $X_0$.
  
  We now move the items in $X_2' \setminus I$ to 0 and consider the allocation 
  \[         (X_0 \cup (X_2' \setminus I), X_1', I). \]
  Observe that $2$ with $I$ does not strongly envy $X_1'$ since it did not strongly envy $X_1' \cup (X_2' \setminus I)$ and $1$ with $X_1'$ does not strongly envy $I$ since it did not strongly envy $X_2'$. But, $X_0 \cup (X_2' \setminus I)$ is a proper superset of $X_0$ and hence infeasible. Thus the above cannot be a completion of $X_0 \cup (X_2' \setminus I)$, and hence either $2$ with $I$ or $1$ with $X_1'$ strongly envies $X_0 \cup (X_2' \setminus I)$. Thus, either $I$ or $X_1'$ is feasible by Lemma~\ref{Feasibility Criterion}, a contradiction. 
\end{proof}

\section{Proof Two}\label{proof2} 
We extend the proof in~\cite{AACGMM22} to compatible pairs. As above, we assume that agents 1 and 2 form a compatible pair. 

\subsection{Repair Tools}
A bundle $B$ in a partition is \emph{EFX-feasible} for agent $i$ if it dominates every one-deletion of every part, i.e., $v_i(B) \ge \rho_i(C)$ for every part $C$. We use the following standard repair due to Plaut and Roughgarden~\cite{PR20}. Let $P$ be any partition of a set of items and let $v$ be a monotone valuation function. As long as there is a bundle $H$ in $P$ and a good $g \in H$ such that $v(H \setminus g) > v(L)$, where $L$ is a bundle of minimum value (breaking ties by cardinality), move $g$ from $H$ to $L$. The move does not decrease the minimum value and lexicographically increases the sorted vector of pairs $(v(B),\abs{B})$, where $B$ ranges over the bundles of $P$. The process terminates, and at termination, the value of any one-deletion is at most the value of the minimum bundle. So every part is EFX-feasible. We denote the resulting partition by $\PR_v(P)$. 

\begin{lemma}[PR repair~\cite{PR20,AACGMM22}]\label{lem:pr}
For a monotone valuation $v$ and any partition $P$ of a set of items, every bundle in the partition  $\PR_v(P)$ is EFX-feasible for $v$ and the value of any bundle in $\PR_v(P)$ is at least the minimum value of a bundle in $P$.
\end{lemma}

As in~\cite{AACGMM22}, ties of agent 0 may be removed by a monotone ordinal refinement $u_0$ that strictly orders all subsets while preserving every strict comparison of $v_0$ (e.g., break ties of $v_0$ first by cardinality and then arbitrarily). Thus every EFX allocation for $(u_0,v_1,v_2)$ is EFX for $(v_0,v_1,v_2)$. Importantly, we do \emph{not} perturb $v_1,v_2$, since doing so could destroy compatibility.

\subsection{Existence proof}
We now insert Lemma~\ref{lem:two-cutter} into the proof architecture of~\cite[Section~4]{AACGMM22}. We give the full argument needed to identify the modified case, while compressing unchanged bookkeeping.

Work with the strict refinement $u_0$ of agent 0, i.e., in the following, agent 0 has valuation $u_0$. Maintain a partition
$\calY=(Y_0,Y_1,Y_2)$ of $M$ such that $Y_0,Y_1$ are EFX-feasible for agent 0 and $Y_2$ is EFX-feasible for at least one of agents 1 and 2 (the \emph{invariant}). Such a partition is initialized by applying $\PR_{u_0}$ to any partition of $M$ into three parts and letting agent 1 choose a favorite part as $Y_2$. Relabel the remaining two so that $u_0(Y_0)<u_0(Y_1)$
and define the potential of the partition as $\phi(\calY)=u_0(Y_0)$.
If the three parts already admit an EFX assignment, stop. Otherwise, we call the partition \emph{blocked}.

\begin{lemma}[Blocked structure]\label{lem:blocked}
At a blocked partition, neither agent 1 nor agent 2 finds either $Y_0$ or $Y_1$ EFX-feasible. Moreover,  $\rho_i(Y_2) > \max\{v_i(Y_0),v_i(Y_1)\}$ for $i\in\{1,2\}$.
\end{lemma}
\begin{proof}
If one of the agents, 1 or 2, finds $Y_0$ or $Y_1$ EFX-feasible, we can assign agents 1 and 2 distinct sets, one of them $Y_2$, which they find EFX-feasible. We assign the remaining set in $Y_0$ or $Y_1$ to agent 0 and obtain an EFX allocation. If $\rho_i(Y_2) \le \max\{v_i(Y_0), v_i(Y_1)\}$ for some $i \in \{1,2\}$, the more valuable set for $i$ among $Y_0$ and $Y_1$ would dominate all one-deletions of $Y_2$ and, by monotonicity, all one-deletions of the less valuable, making it EFX-feasible.
\end{proof}

We repeatedly use the following \emph{restoration fact} from~\cite{AACGMM22}. If in a partition $(A,B,C)$ of $M$, $A$ and $B$ are mutually EFX for agent 0 (i.e., agent 0 with $A$ is EFX toward $B$ and with $B$ is EFX toward $A$), $C$ is EFX-feasible for agent 1 or 2, and
\begin{equation}\label{eq:threshold}
 \min\{u_0(A),u_0(B)\}>\phi(\calY),
\end{equation}
then either $A$ and $B$ are already EFX-feasible for agent 0, or applying $\PR_{u_0}$ as in the initialization restores the invariant; in both cases the new potential is strictly larger than $\phi(\calY)$. This follows because PR never decreases the minimum bundle value.

For $i \in \{1,2\}$, let $g_i$ be an item of smallest marginal value in $Y_2$, i.e., $v_i(Y_2 \setminus g_i) = \rho_i(Y_2)$. There are two cases.

\paragraph{Case I: Residual Dominance.}
Suppose for some $i\in\{1,2\}$,
\begin{equation}\label{eq:case1}
 v_i(Y_2 \setminus g_i)\ge v_i(Y_0 \cup g_i).
\end{equation}
Then, by Lemma~\ref{lem:blocked}, $Y_2 \setminus g_i$ is a favorite part of $(Y_0 \cup g_i,Y_1,Y_2 \setminus g_i)$ for this $i$. Choose an inclusion-minimal $T\subseteq Y_0 \cup g_i$ with $u_0(T)>u_0(Y_0)$, and form
\[
 A=T,\quad B=Y_1,\quad
 C=(Y_2 \setminus g_i)\cup((Y_0 \cup g_i)\setminus T).
\]
The part $C$ remains a favorite for agent $i$. Minimality of $T$ gives $u_0(T \setminus g)\le u_0(Y_0)$ for every $g\in T$, while old EFX-feasibility of $Y_0$ gives $u_0(Y_0)\ge u_0(Y_1 \setminus h)$ for every $h\in Y_1$. Hence $T,Y_1$ are mutually EFX for agent 0, and
$\min\{u_0(T),u_0(Y_1)\}>u_0(Y_0)=\phi(\calY)$. Restoration strictly increases the potential. This case is unchanged from~\cite{AACGMM22} and uses no compatibility assumption.

\paragraph{Case II: The Modified Case.}
Assume instead
\begin{equation}\label{eq:case2}
 v_i(Y_2 \setminus g_i)<v_i(Y_0 \cup g_i),\qquad \text{for } i \in \{1,2\}.
\end{equation}
Let $S=Y_0\cup Y_2$ and define
\[
 P^{(1)}=(Y_0 \cup g_1,Y_2 \setminus g_1),\qquad
 P^{(2)}=(Y_0 \cup g_2,Y_2 \setminus g_2).
\]
Define $t_i = \low_i(P^{(i)})$ for $i \in \{1,2\}$. Then, by Lemma~\ref{lem:blocked} and \eqref{eq:case2},
\begin{equation}\label{eq:targets}
 t_i =v_i(Y_2 \setminus g_i)>v_i(Y_1),\qquad \text{for } i \in \{1,2\}.
\end{equation}
We now apply Lemma~\ref{lem:two-cutter} on $S$. It returns an EFX allocation $(D_1,D_2)$ of $S$ between agents 1 and 2 with $v_i(D_i)\ge t_i$. Therefore,
\[
 (Y_1,D_1,D_2)
\]
is EFX for agents 1 and 2; indeed, for every $h\in Y_1$ we have
$v_i(D_i)\ge t_i>v_i(Y_1)\ge v_i(Y_1 \setminus h)$.
Notice that the untouched bundle is $Y_1$ not $Y_0$, because the application of Lemma~\ref{lem:two-cutter} repartitions all of $Y_0\cup Y_2$.

If $Y_1$ is EFX-feasible for agent 0 in $(Y_1,D_1,D_2)$, we are done. Otherwise, agent 0 strongly envies some $D_p$, $p\in\{1,2\}$. Choose an inclusion-minimal $T\subseteq D_p$ satisfying $u_0(T)>u_0(Y_1)$, and let $q$ denote the other agent. Set
\[
 A=Y_1,\qquad B=T,\qquad
 C=D_q\cup(D_p\setminus T).
\]
Minimality of $T$ gives $u_0(T \setminus g)\le u_0(Y_1)$ for every $g\in T$, so $Y_1,T$ are mutually EFX for agent 0. Agent $q$ finds $C$ EFX-feasible: $C$ contains her old bundle $D_q$, whereas the two opposing bundles are $Y_1$ and a subset $T\subseteq D_p$, and $D_q$ was EFX-feasible for $q$ in $(Y_1,D_1,D_2)$. Finally,
\[
 \min\{u_0(Y_1),u_0(T)\}=u_0(Y_1)>u_0(Y_0)=\phi(\calY).
\]
Restoration therefore yields a partition with strictly larger potential.

\begin{proof}[Proof of Theorem~\ref{thm:main}]
At every blocked partition, Lemma~\ref{lem:blocked} applies and exactly one of Cases I and II occurs. Each case either outputs an EFX allocation or restores the invariant with strictly larger potential. Since $\phi$ is the $u_0$-value of a subset of $M$, it has only finitely many possible values. Hence the process terminates.  Finally, EFX for the refined valuation $u_0$ implies EFX for the original $v_0$.
\end{proof}

\subsection{Methods and Disclosure}\label{methods and disclosure}
Proof One of Theorem~\ref{thm:main} (Section~\ref{proof1}) was discovered by AI systems directed by the human authors, as a by-product of an attempt to prove EFX for four additive agents. ChatGPT (OpenAI, model ChatGPT.6~Pro) proved the central step, the repair in Lemma~\ref{construction of a witness}, and found Lemma~\ref{lem:two-cutter}, which extends the proof to compatible pairs. Claude (Anthropic, model Claude Fable~5.1) checked these proofs. Proof Two (Section~\ref{proof2}) was found by the human authors. All proofs were checked by the human authors.

\subsection{Discussion}
The proof changes only one point of~\cite{AACGMM22}. Their case II uses MMS-feasibility to guarantee the second agent's target after a PR repair performed according to another valuation. We use Lemma~\ref{lem:two-cutter}. 

Each iteration of the algorithm, except the last, which gives the EFX allocation, improves the potential $\phi(\calY) = \min\{u_0(Y_0), u_0(Y_1)\}$. Starting with a partition that satisfies the invariant and maximizes the potential removes the iteration from the second proof and constructs the EFX allocation with one modification, similarly to the first proof.

\bibliographystyle{alpha}
\bibliography{ref}

@misc{AACGMM22,
  author = {Akrami, Hannaneh and Alon, Noga and Chaudhury, Bhaskar Ray and Garg, Jugal and Mehlhorn, Kurt and Mehta, Ruta},
  title = {{EFX} Allocations: Simplifications and Improvements},
  year = {2022},
  howpublished = {arXiv:2205.07638v2},
  note = {Journal version: {EFX}: A Simpler Approach and an (Almost) Optimal Guarantee via Rainbow Cycle Number, \emph{Operations Research}, 73(2):738--751, 2025},
  url = {https://arxiv.org/abs/2205.07638v2}
}

@article{PR20,
  author = {Plaut, Benjamin and Roughgarden, Tim},
  title = {Almost Envy-Freeness with General Valuations},
  journal = {SIAM Journal on Discrete Mathematics},
  year = {2020},
  volume = {34},
  number = {2},
  pages = {1039--1068},
  doi = {10.1137/19M124397X},
  url = {https://doi.org/10.1137/19M124397X}
}

@article{CGM24,
  author = {Chaudhury, Bhaskar Ray and Garg, Jugal and Mehlhorn, Kurt},
  title = {{EFX} Exists for Three Agents},
  journal = {Journal of the ACM},
  year = {2024},
  volume = {71},
  number = {1},
  pages = {4:1--4:27},
  doi = {10.1145/3616009},
  url = {https://doi.org/10.1145/3616009},
  note = {Preliminary version at EC 2020; arXiv:2002.05119}
}

@inproceedings{BCFF22,
  author = {Berger, Ben and Cohen, Avi and Feldman, Michal and Fiat, Amos},
  title = {Almost Full {EFX} Exists for Four Agents},
  booktitle = {Proceedings of the 36th AAAI Conference on Artificial Intelligence (AAAI)},
  year = {2022},
  volume = {36},
  pages = {4826--4833},
  doi = {10.1609/aaai.v36i5.20410},
  note = {Full version: arXiv:2102.10654}
}

@article{Mahara24,
  author = {Mahara, Ryoga},
  title = {Extension of Additive Valuations to General Valuations on the Existence of {EFX}},
  journal = {Mathematics of Operations Research},
  year = {2024},
  volume = {49},
  number = {2},
  pages = {1263--1277},
  doi = {10.1287/moor.2022.0044},
  url = {https://doi.org/10.1287/moor.2022.0044},
  note = {Preliminary version at ESA 2021; arXiv:2107.09901}
}

@inproceedings{AGS25,
  author = {Ashuri, Arash and Gkatzelis, Vasilis and Sgouritsa, Alkmini},
  title = {{EF2X} Exists for Four Agents},
  booktitle = {Proceedings of the 39th AAAI Conference on Artificial Intelligence (AAAI)},
  year = {2025},
  volume = {39},
  pages = {13555--13563},
  doi = {10.1609/aaai.v39i13.33480}
}

@misc{AACMS26,
  author = {Akrami, Hannaneh and Ashuri, Arash and Chaudhury, Bhaskar Ray and Mehlhorn, Kurt and Soldatov, Maksim},
  title = {A Simple Polynomial-Time {EFX} Repair for Cancelable Valuations},
  year = {2026},
  howpublished = {arXiv:2608.08864},
  url = {https://arxiv.org/abs/2608.08864}
}

@article{CKMPSW19,
  author = {Caragiannis, Ioannis and Kurokawa, David and Moulin, Herv{\'e} and Procaccia, Ariel D. and Shah, Nisarg and Wang, Junxing},
  title = {The Unreasonable Fairness of Maximum {N}ash Welfare},
  journal = {ACM Transactions on Economics and Computation},
  year = {2019},
  volume = {7},
  number = {3},
  pages = {12:1--12:32},
  doi = {10.1145/3355902},
  note = {Preliminary version at EC 2016}
}

@inproceedings{GMT14,
  author = {Gourv{\`e}s, Laurent and Monnot, J{\'e}r{\^o}me and Tlilane, Lydia},
  title = {Near Fairness in Matroids},
  booktitle = {Proceedings of the 21st European Conference on Artificial Intelligence (ECAI)},
  series = {Frontiers in Artificial Intelligence and Applications},
  volume = {263},
  pages = {393--398},
  publisher = {IOS Press},
  year = {2014},
  doi = {10.3233/978-1-61499-419-0-393}
}

@article{Bud11,
  author = {Budish, Eric},
  title = {The Combinatorial Assignment Problem: Approximate Competitive Equilibrium from Equal Incomes},
  journal = {Journal of Political Economy},
  year = {2011},
  volume = {119},
  number = {6},
  pages = {1061--1103}
}

@inproceedings{LMMS04,
  author = {Lipton, Richard J. and Markakis, Evangelos and Mossel, Elchanan and Saberi, Amin},
  title = {On Approximately Fair Allocations of Indivisible Goods},
  booktitle = {Proceedings of the 5th ACM Conference on Electronic Commerce (EC)},
  year = {2004},
  pages = {125--131}
}

@article{CKMS21,
  author = {Chaudhury, Bhaskar Ray and Kavitha, Telikepalli and Mehlhorn, Kurt and Sgouritsa, Alkmini},
  title = {A Little Charity Guarantees Almost Envy-Freeness},
  journal = {SIAM Journal on Computing},
  year = {2021},
  volume = {50},
  number = {4},
  pages = {1336--1358},
  note = {Preliminary version at SODA 2020}
}

@article{AMN20,
  author = {Amanatidis, Georgios and Markakis, Evangelos and Ntokos, Apostolos},
  title = {Multiple Birds with One Stone: Beating 1/2 for {EFX} and {GMMS} via Envy Cycle Elimination},
  journal = {Theoretical Computer Science},
  year = {2020},
  volume = {841},
  pages = {94--109},
  note = {Preliminary version at AAAI 2020}
}

@article{ABFHV21,
  author = {Amanatidis, Georgios and Birmpas, Georgios and Filos-Ratsikas, Aris and Hollender, Alexandros and Voudouris, Alexandros A.},
  title = {Maximum {N}ash Welfare and Other Stories About {EFX}},
  journal = {Theoretical Computer Science},
  year = {2021},
  volume = {863},
  pages = {69--85},
  doi = {10.1016/j.tcs.2021.02.020},
  note = {Preliminary version at IJCAI 2020}
}

@inproceedings{HGNV25,
  author = {Vishwa Prakash {H.V.} and Ghosal, Pratik and Nimbhorkar, Prajakta and Varma, Nithin},
  title = {{EFX} Exists for Three Types of Agents},
  booktitle = {Proceedings of the 26th ACM Conference on Economics and Computation (EC)},
  year = {2025},
  pages = {101--128},
  doi = {10.1145/3736252.3742509}
}

@misc{AFM26,
  author = {Alkassar, Eyad and Fouz, Mahmoud and Mehlhorn, Kurt},
  title = {Complete {EFX} Allocations Exist for Four Additive Agents and Up to Nine Goods},
  year = {2026},
  howpublished = {arXiv:2608.08590},
  url = {https://arxiv.org/abs/2608.08590}
}

@inproceedings{CFKS23,
  author = {Christodoulou, George and Fiat, Amos and Koutsoupias, Elias and Sgouritsa, Alkmini},
  title = {Fair Allocation in Graphs},
  booktitle = {Proceedings of the 24th ACM Conference on Economics and Computation (EC)},
  year = {2023},
  pages = {473--488}
}

@misc{AAMMS26,
  author = {Afshinmehr, Mahyar and Ashuri, Arash and Mahmoudkhan, Pouria and Mehlhorn, Kurt and Shahrezaei, Amir Mohammad},
  title = {{EFX} Allocations Exist on Multi-Graphs},
  year = {2026},
  howpublished = {arXiv:2606.18665},
  url = {https://arxiv.org/abs/2606.18665}
}

@inproceedings{BBMN18,
  author       = {Siddharth Barman and
                  Arpita Biswas and
                  Sanath Kumar Krishna Murthy and
                  Yadati Narahari},
  title        = {Groupwise Maximin Fair Allocation of Indivisible Goods},
  booktitle    = {Proceedings of the Thirty-Second {AAAI} Conference on Artificial Intelligence},
  pages        = {917--924},
  publisher    = {{AAAI} Press},
  year         = {2018},
  doi          = {10.1609/AAAI.V32I1.11463},
}

@article{GHH25,
  author = {Goldberg, Paul W. and H{\o}gh, Kasper and Hollender, Alexandros},
  title = {The Frontier of Intractability for {EFX} with Two Agents},
  journal = {Theoretical Computer Science},
  year = {2025},
  pages = {115367},
  doi = {10.1016/j.tcs.2025.115367},
  note = {Preliminary version at SAGT 2023}
}

@inproceedings{AAMM26,
  author       = {Mahyar Afshinmehr and
                  Arash Ashuri and
                  Pouria Mahmoudkhan and
                  Kurt Mehlhorn},
  title        = {{EFX} Allocations Exist on Triangle-Free Multi-Graphs},
  booktitle    = {Proceedings of the 25th International Conference on Autonomous Agents
                  and Multiagent Systems, {AAMAS}},
  pages        = {2419--2428},
  publisher    = {{ACM}},
  year         = {2026},
  doi          = {10.65109/XYOA9688},
}

@inproceedings{CGRSV22,
  author       = {Ioannis Caragiannis and
                  Jugal Garg and
                  Nidhi Rathi and
                  Eklavya Sharma and
                  Giovanna Varricchio},
  title        = {New Fairness Concepts for Allocating Indivisible Items},
  booktitle    = {Proceedings of the Thirty-Second International Joint Conference on
                  Artificial Intelligence, {IJCAI}},
  pages        = {2554--2562},
  publisher    = {ijcai.org},
  year         = {2023},
  doi          = {10.24963/IJCAI.2023/284},
}

@inproceedings{HANR25,
  author       = {Hannaneh Akrami and
                  Nidhi Rathi},
  title        = {Epistemic {EFX} Allocations Exist for Monotone Valuations},
  booktitle    = {AAAI-25, Sponsored by the Association for the Advancement of Artificial
                  Intelligence, February 25 - March 4, 2025, Philadelphia, PA, {USA}},
  publisher    = {{AAAI} Press},
  year         = {2025},
}

@inproceedings{AG25,
  author       = {Arash Ashuri and
                  Vasilis Gkatzelis},
  title        = {Simultaneously Satisfying {MXS} and {EFL}},
  booktitle    = {Proceedings of the 26th {ACM} Conference on Economics and Computation,
                  {EC} 2025, Stanford University, Stanford, CA, USA, July 7-10, 2025},
  pages        = {689--718},
  publisher    = {{ACM}},
  year         = {2025},
}

@inproceedings{AR25,
  author       = {Hannaneh Akrami and
                  Nidhi Rathi},
  title        = {Achieving Maximin Share and {EFX/EF1} Guarantees Simultaneously},
  booktitle    = {AAAI-25, Sponsored by the Association for the Advancement of Artificial
                  Intelligence, February 25 - March 4, 2025, Philadelphia, PA, {USA}},
  pages        = {13529--13537},
  publisher    = {{AAAI} Press},
  year         = {2025},
}

@misc{AFMMR26,
      title={The Power of Share-Based Notions in Proving Envy-Based Fairness Guarantees}, 
      author={Hannaneh Akrami and Uriel Feige and Ryoga Mahara and Kurt Mehlhorn and Nidhi Rathi},
      year={2026},
      eprint={2602.11732},
      archivePrefix={arXiv},
      primaryClass={cs.GT},
      url={https://arxiv.org/abs/2602.11732}, 
}

@misc{AAM26,
      title={Polynomial Algorithms for Simultaneous Share-Based and Envy-Based Guarantees}, 
      author={Hannaneh Akrami and Arash Ashuri and Kurt Mehlhorn},
      year={2026},
      eprint={2610.15437},
      archivePrefix={arXiv},
      primaryClass={cs.GT},
      url={https://arxiv.org/abs/2610.15437}, 
}

@misc{AMMMSSW26,
      title={EFX Need Not Exist}, 
      author={Hannaneh Akrami and Simon Mackenzie and Alexander Mayorov and Kurt Mehlhorn and Shreyas Srinivas and Mashbat Suzuki and Christoph Weidenbach},
      year={2026},
      eprint={2604.18216},
      archivePrefix={arXiv},
      primaryClass={cs.GT},
      url={https://arxiv.org/abs/2604.18216}, 
}

\end{document}